\documentclass[aps,prl,twocolumn,unsortedaddress,superscriptaddress,nofootinbib,10pt]{revtex4-2}
\usepackage{graphicx,amsmath,amssymb,comment}
\usepackage[utf8x]{inputenc}
\usepackage{physics}
\usepackage{mathrsfs}
\usepackage{todonotes}
\usepackage{bm}
\usepackage{graphicx}
\usepackage{hyperref}
\date{\today}
\begin{document}
%\linenumbers
\title{Storage and retrieval of optical skyrmions with topological characteristics}

\author{Jinwen Wang}
\email{jinwenwang@xjtu.edu.cn}
% \email{J. W. and X. Y. contributed equally to this work.}
\affiliation{Ministry of Education Key Laboratory for Nonequilibrium Synthesis and Modulation of Condensed Matter, Shaanxi Province Key Laboratory of Quantum Information and Quantum Optoelectronic Devices, School of Physics, Xi'an Jiaotong University, Xi'an 710049, China}
% \altaffiliation{J. W. and X. Y. contributed equally to this work.}

\author{Xin Yang}
\email{J. W. and X. Y. contributed equally to this work.}
\affiliation{Ministry of Education Key Laboratory for Nonequilibrium Synthesis and Modulation of Condensed Matter, Shaanxi Province Key Laboratory of Quantum Information and Quantum Optoelectronic Devices, School of Physics, Xi'an Jiaotong University, Xi'an 710049, China}

\author{Yun Chen}
\affiliation{School of Science, Huzhou University, Zhejiang 313000, China}

\author{Zhujun Ye}
\affiliation{International Institute for Sustainability with Knotted Chiral Meta Matter (WPI-SKCM$^2$), Hiroshima University, 1-3-1 Kagamiyama, Higashi-Hiroshima, Hiroshima 739-8526, Japan}

\author{Xinji Zeng}
\affiliation{Ministry of Education Key Laboratory for Nonequilibrium Synthesis and Modulation of Condensed Matter, Shaanxi Province Key Laboratory of Quantum Information and Quantum Optoelectronic Devices, School of Physics, Xi'an Jiaotong University, Xi'an 710049, China}

\author{Yongkun Zhou}
\affiliation{Ministry of Education Key Laboratory for Nonequilibrium Synthesis and Modulation of Condensed Matter, Shaanxi Province Key Laboratory of Quantum Information and Quantum Optoelectronic Devices, School of Physics, Xi'an Jiaotong University, Xi'an 710049, China}

\author{Shuya Zhang}
\affiliation{Ministry of Education Key Laboratory for Nonequilibrium Synthesis and Modulation of Condensed Matter, Shaanxi Province Key Laboratory of Quantum Information and Quantum Optoelectronic Devices, School of Physics, Xi'an Jiaotong University, Xi'an 710049, China}

\author{Claire Marie Cisowski}
\affiliation{School of Physics and Astronomy, University of Glasgow, G12 8QQ, United Kingdom}

\author{Chengyuan Wang}%
\affiliation{Ministry of Education Key Laboratory for Nonequilibrium Synthesis and Modulation of Condensed Matter, Shaanxi Province Key Laboratory of Quantum Information and Quantum Optoelectronic Devices, School of Physics, Xi'an Jiaotong University, Xi'an 710049, China}

\author{Katsuya Inoue}
\affiliation{International Institute for Sustainability with Knotted Chiral Meta Matter (WPI-SKCM$^2$), Hiroshima University, 1-3-1 Kagamiyama, Higashi-Hiroshima, Hiroshima 739-8526, Japan}

\author{Yijie Shen}
\email{yijie.shen@ntu.edu.sg}
\affiliation{Centre for Disruptive Photonic Technologies, School of Physical and Mathematical Sciences, Nanyang Technological University, Singapore 637371, Singapore}
\affiliation{School of Electrical and Electronic Engineering, Nanyang Technological University, Singapore 639798, Republic of Singapore}
\affiliation{International Institute for Sustainability with Knotted Chiral Meta Matter (WPI-SKCM$^2$), Hiroshima University, 1-3-1 Kagamiyama, Higashi-Hiroshima, Hiroshima 739-8526, Japan}

\author{Sonja Franke-Arnold}
\email{sonja.franke-arnold@glasgow.ac.uk}
\affiliation{School of Physics and Astronomy, University of Glasgow, G12 8QQ, United Kingdom}

\author{Hong Gao}
\email{honggao@xjtu.edu.cn}
\affiliation{Ministry of Education Key Laboratory for Nonequilibrium Synthesis and Modulation of Condensed Matter, Shaanxi Province Key Laboratory of Quantum Information and Quantum Optoelectronic Devices, School of Physics, Xi'an Jiaotong University, Xi'an 710049, China}

\begin{abstract}
Optical skyrmions are topological structures of light whose defining property, the skyrmion number, is robust against perturbations. This makes them attractive for applications in quantum information storage, where resilience to decoherence is paramount. However, their preservation during coherent storage remains unexplored. We report the first experimental demonstration of storing and retrieving optical skyrmions in a cold $^{87}$Rb vapor using a dual-path electromagnetically induced transparency memory. Crucially, we show that the skyrmion number remains invariant for storage times up to several microseconds, even when subjected to imbalanced loss between the two paths and substantial perturbations in control beam power. Our work demonstrates the survival of a non-trivial topological invariant in a quantum memory, marking a significant step towards topologically protected photonic technologies.

\end{abstract}

\maketitle
\paragraph{\textbf{Introduction:}}
Optical skyrmions, as topological quasiparticles with distinct polarization textures, have become an active area of research in recent years \cite{shen2024optical,yang2025optical,ALLAM2025}. These structures promise inherent stability against perturbations, making them potential candidates for photonic devices and information processing technologies \cite{romming2013writing,wiesendanger2016nanoscale,porfirev2023light}. Originally proposed in the context of particle physics, the concept of skyrmions has since then been translated to various physical systems, such as magnetic materials \cite{mühlbauer2009skyrmion,bogdanov2020physical}, ultra-cold atoms \cite{al2001skyrmions}, acoustic systems \cite{ge2021observation}, and fluid systems \cite{smirnova2024water,wang2025topological}. In optics, skyrmions manifest as vectorial textures of light fields, allowing the exploration of various skyrmionic structures \cite{tsesses2018optical,du2019deep,gao2020paraxial,karnieli2021emulating,zhang2021bloch,shen2021topological,cisowski2023building,mcwilliam2023topological,sugic2021particle,ehrmanntraut2023optical,shen2023topological}. Their preparation is comparatively straightforward, since optical systems are free from the crystal structure constraints inherent to condensed matter systems. Recent studies of optical skyrmions demonstrate that their topological information could resist noise and defects in complex systems \cite{liu2022disorder,wang2024topological1,wang2024topological2}, with potential applications in light-matter interactions \cite{wang2020vectorial,wu2022conformal,mitra2025topological,castellucci2021atomic,jia2025electrically}, as well as classical or quantum communication and computation \cite{he2024optical,ornelas2024non,de2025quantum,ma2025nanophotonic,wang2025perturbation}. For instance, quantum information could be encoded in a two-photon entangled state defined by a fixed skyrmion number, where the topological protection suppresses noise effects and keeps the global invariant \cite{ornelas2025topological}. 

Quantum networks with topological quasiparticles require quantum memories in addition to communication channels. So far, the optical storage of skyrmionic beams remains unexplored. Storing polarization-encoded information, such as polarization qubits \cite{namazi2017ultralow,vernaz2018highly,wang2019efficient} and vector beams \cite{parigi2015storage,ye2019experimental,zeng2023optical,yang2025efficient} in atomic ensembles typically requires a dual-path interferometric configuration to cancel polarization sensitivity, which arises due to the presence of multiple Zeeman sublevels. This approach, however, demands highly balanced efficiency and phase stability between the two paths, as any imperfection severely degrades the output fidelity \cite{sangouard2011quantum}.
%In the context of optical and quantum storage systems utilizing polarization encoding, several notable advances have been made in recent years. Researchers have successfully demonstrated the storage of structured light carrying polarization information, including polarization qubits \cite{namazi2017ultralow,vernaz2018highly,wang2019efficient} and vector beams \cite{parigi2015storage,ye2019experimental,zeng2023optical,yang2025efficient}, within atomic ensembles. The widely adopted strategy to mitigate the inherent polarization sensitivity of polarized atomic media, which arises due to the presence of multiple Zeeman sublevels, is the conversion of polarization-encoded information into path-encoded information via a dual-path interferometric configuration. Although this strategy enables storage of polarization states, it introduces strict experimental demands: both optical paths must exhibit highly balanced storage efficiency and exceptional phase stability throughout the process. Any imperfections between two paths can significantly degrade output fidelity and impair polarization reconstruction.
In light of these challenges, optical skyrmions emerge as a promising alternative information carrier that could overcome the limitations of conventional polarization storage schemes. Their topological properties could be preserved under unitary or non-unitary transformations \cite{zhang2025topological,guo2025topological,he2025reconfigurable,peters2025seeing} which are commonly encountered in optical systems, for instance in scattering or propagation through noisy channels, making them resistant to decoherence and mode distortion. This suggested topological protection could provide a reliable stability that potentially cancels the effective noise arising from decoherence, efficiency imbalances, and phase differences between the two paths, and still retains a global topological invariant and a robust quantum number. 

%Current optical storage technologies rely mainly on encoding schemes based on intensity, polarization, time-binning, or spatial modes, namely orbital angular momentum (OAM) \cite{zhao2009millisecond,hosseini2011unconditional,wang2019efficient,clausen2011quantum,liu2025millisecond,dong2023highly}. However, these approaches have inherent limitations in noise resilience and storage capacity. Polarization-based encoding, for example, is susceptible to the anisotropic response induced by the storage medium or upon free-space transmission. High-order OAM modes are prone to pronounced distortion and decomposition into single vortices, leading to reduced storage fidelity. Since these structured modes are essential for high-dimensional quantum communication, such imperfections can compromise the robustness of quantum networks against system perturbations. In contrast, the skyrmion number, as a topological quantum number, may offer inherent advantages in condensed matter systems \cite{nagaosa2013topological,fert2017magnetic,zhang2020skyrmion} and for optical realizations \cite{Nape2022}.

\begin{figure*}[!t]
\centerline{\includegraphics[width=0.95\linewidth]{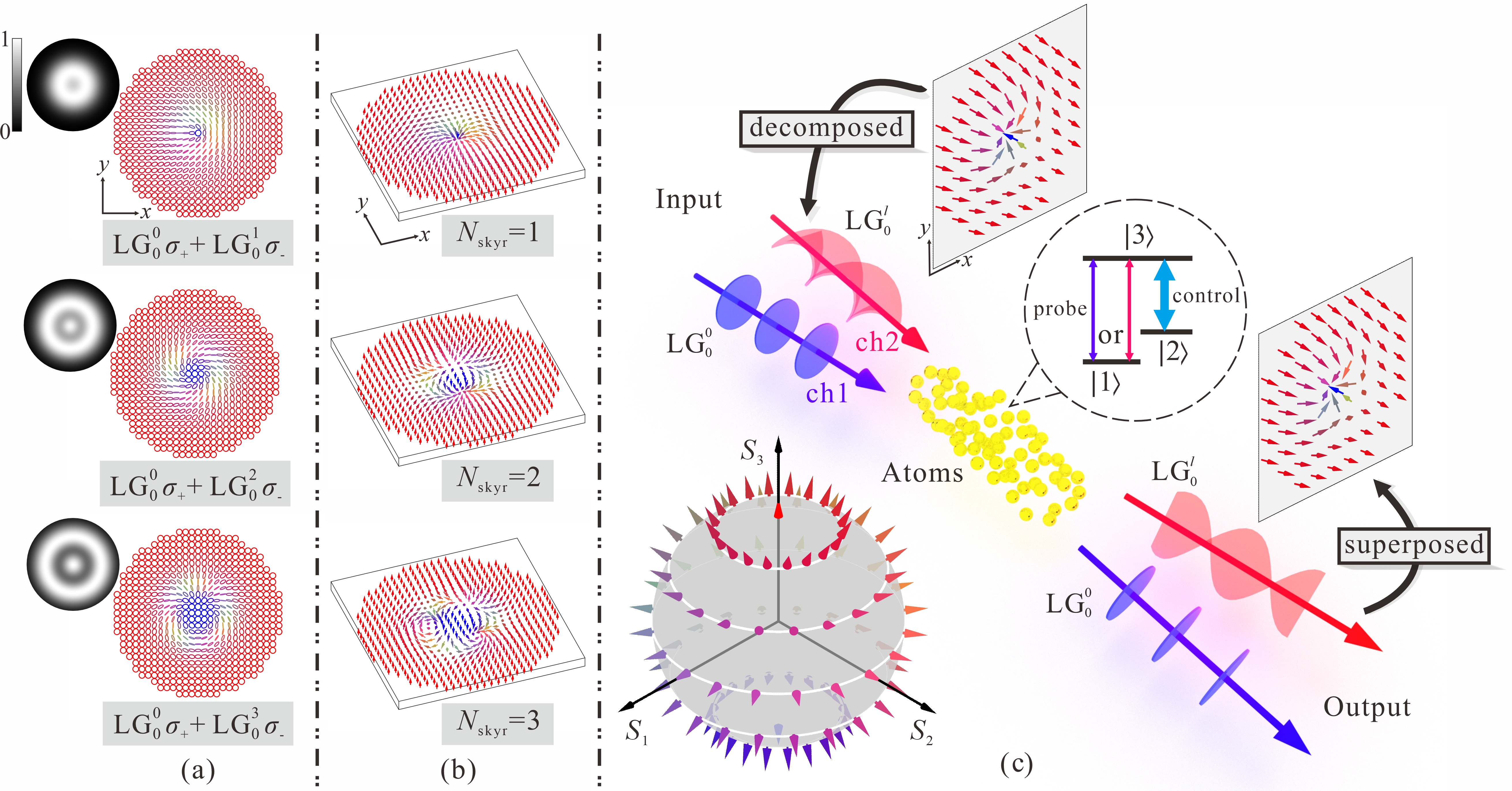}}
\caption{(a) Simulated polarization and intensity distributions of paraxial optical skyrmions for ${N}_\mathrm{skyr}=1$, 2, 3 in Eq.~(\ref{eq1}). (b) 3D perspective of simulated Stokes vectors for optical skyrmions with the corresponding ${N}_{\rm{skyr}}$ values. (c) The schematic diagram of optical storage. The probe beam drives the level $|1\rangle$ to $|3\rangle$ in each path, and the control beam simultaneously drives the level $|2\rangle$ to $|3\rangle$ in both paths. We input the signals that need to be stored in the atomic storage medium, and obtain the output (retrieved) signals after storage. More details are in the main text. The left bottom inset shows the Stokes vector $\pmb{s}$ on the unit Poincar\'{e} sphere. And the mapping relationship between arrows and polarizations are indicated by the same color.}
\label{fig1}
\end{figure*}

In this work, we realize the optical storage of optical skyrmions by employing a dual-path storage scheme based on an electromagnetically induced transparency (EIT) protocol. The two spatial modes of the optical skyrmion are spatially separated and independently stored within cold $^{87}$Rb atoms. 
Differences in the intensity distribution of the two spatial mode profiles lead to an unavoidable imbalance in storage efficiency and a phase difference due to dispersion accumulated along the two paths. In addition, the system experiences a further decoherence-induced loss of storage efficiency. Our experimental results demonstrate, that nevertheless the skyrmion number is conserved for a storage time of several microseconds.
%An imbalance in storage efficiency and a phase change naturally occur between the two paths due to differences in the spatial profiles of the modes, resulting in varying interaction strengths. However, the experimental results demonstrate that the skyrmion number remains robust for a storage time of a few microseconds, despite the system experiencing a further decoherence-induced loss of storage efficiency. 
While EIT-based storage is well-established, here, we demonstrate for the first time that the topological invariant of an optical skyrmion beam remains robust, which is not achievable with conventional polarization or orbital angular momentum (OAM) encoding. The results of our study provide not only experimental evidence for the stability of a topologically non-trivial photonic encoding space against realistic memory imperfections but also offer valuable insights into future storage devices based on skyrmionic structures \cite{yu2017room,fert2017magnetic,zhang2020skyrmion}.

\paragraph{\textbf{Concept and model:}}
To study the storage of optical skyrmions, we begin by generating states of the form
\begin{align}
&|\Psi^{\it{l}}\rangle = \frac{1}{\sqrt{2}}\left(|\rm{LG}^{0}_{0}\rangle|{\boldsymbol\sigma}_+\rangle + |\rm{LG}^{\it{l}}_{0}\rangle|{\boldsymbol\sigma}_-\rangle\right),
\label{eq1}
\end{align}
where $p$ (the lower index) and $l$ denote the radial and azimuthal mode number of Laguerre-Gaussian (LG) modes \cite{franke2017optical}, and $\boldsymbol\sigma_+$, $\boldsymbol\sigma_-$ are left and right circular polarizations. 
%This combination ensures opposite circular polarization states at the center and the periphery, along with a continuous transition through elliptical and linear polarization states. 
Thus, the polarization structure depends on the relative local intensity and phase between the two modes, and thereby on the azimuthal quantum number $l$ of the LG mode \cite{ye2024theory}. The simulated polarization distributions and Stokes vectors of paraxial optical skyrmions with corresponding skyrmion numbers ${N}_\mathrm{skyr}=1$, 2, and 3 are shown in Figs.~\ref{fig1}(a) and (b). %Here, the Gaussian mode $\rm{LG}^{{0}}_{0}$ is a common component in the superposition for all generated skyrmions. 
More details about calculation of optical skyrmions can be found in the Supplemental Material.

\begin{figure*}[!t]
\centerline{\includegraphics[width=\linewidth]{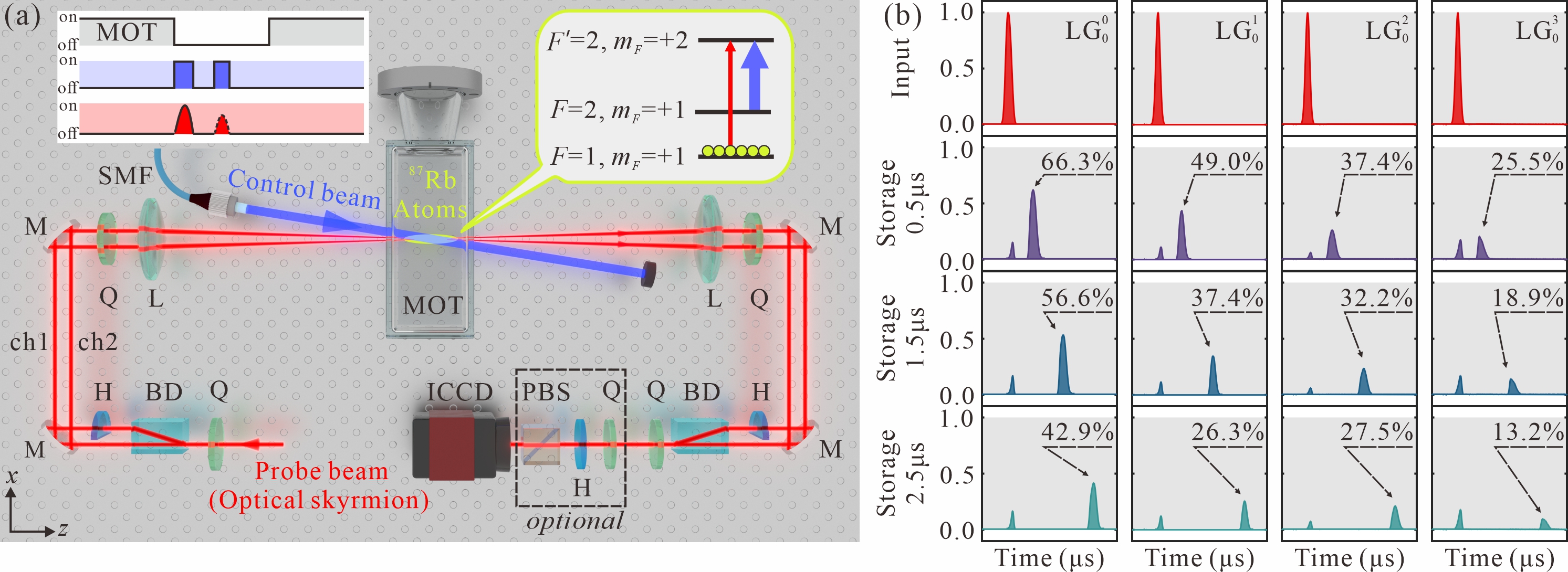}}
\caption{(a) Experimental setup for the optical skyrmion storage. The control beam, with a waist size of 4 mm and a power of 33 mW, and is incident on the atomic medium at a $1^\circ$ angular separation relative to probe beams. By controlling the timing sequence of the MOT and optical switching (top left inset), as well as locking the frequency and phase of two beams (top right inset), the storage and retrieving processes are achieved. Full spatially resolved Stokes tomography (optional) is performed to characterize the polarization structure. Optical elements: H, half-wave plate; L, lens; M, mirror. (b) Temporal waveforms of the input LG beams, which both have Gaussian temporal profiles with 0.5 $\mu$s width, and their corresponding retrieved signals. The storage efficiencies for all paths are presented. The non-stored (smaller) remnants of the pulses are the leakage.}
\label{fig2}
\end{figure*}
%(four times larger than that of the probe beam)
%After storage, the same optical components used in the decomposition step are employed to coherently superimpose the two separated probe beams, thereby reconstructing the optical skyrmion, which is then imaging onto an ICCD.

As illustrated in Fig. \ref{fig1}(c), the input optical skyrmion is converted from a spatial mode and polarization correlated state to a spatial mode and path correlated state using conventional polarization optics. The state is now 
\begin{align}
|\Psi^{\it{l}}\rangle = \frac{1}{\sqrt{2}}\left(\ket{\rm{LG}^{0}_{0}}\ket{\rm{ch_1}} + |\rm{LG}^{\it{l}}_{0}\rangle\ket{\rm{ch_2}}\right){\boldsymbol\sigma}_+,
\label{eq3}
\end{align}
a format that lends itself to dual-path EIT storage. Each spatial mode encoded with ${\boldsymbol\sigma}_+$ polarization to interface with the atomic $\Lambda$-system \cite{hsiao2018highly}, matching the control beam. The associated Poincar\'e sphere picture would feature the two paths as North and South pole, spanning complex superpositions of modal components along the two paths. In each storage path, the $\rm{LG}^{\it{l}}_{0}$ mode with a Rabi frequency $\Omega_l$ drives the level $|1\rangle$ to $|3\rangle$, and one shared control beam simultaneously drives the level $|2\rangle$ to $|3\rangle$. Under the slowly-varying envelope approximation, the Maxwell equation for the probe field is \cite{fleischhauer2005electromagnetically,hsiao2018highly,dong2023highly}:
\begin{align}
\frac{\partial\Omega_{l}}{\partial z}+\frac{1}{c}\frac{\partial\Omega_{l}}{\partial t}=i\frac{D_{l}\Gamma}{2L}\rho_{31},
\label{eq4}
\end{align}
where $\Gamma$ denotes the decay rate of $|3\rangle$, $L$ is the length of the atomic medium, and $\rho_{31}$ represents the atomic coherence between levels $|1\rangle$ and $|3\rangle$. $D_{l}$ is the effective optical depth (OD) of the atomic medium for the $\rm{LG}^{\it{l}}_{0}$ mode. The beam waist of LG modes scales as $\sqrt{l+1}$, leading to a reduced OD and weaker light-atom interaction for higher order modes. It can be observed from Eq. \ref{eq4} that $D_{l}$ affects the storage performance of each path significantly, and consequently causes an imbalance in storage efficiency, acting as an effective noise source. After a specified storage time, the control beam triggers the read-out of these two components, which are recombined so that the topological structures of the (retrieved) output signals can be analyzed. By solving the Maxwell–Bloch equations, we derive the numerical relation between the storage efficiency and the $l$ for each path. We further compare this with the concurrence \cite{wang2024measuring} of vector beams, which are more resistant to perturbations than OAM modes \cite{nape2022revealing}, as detailed in the Supplemental Material.

Although the dual-path storage protocol introduces asymmetric efficiency and hence is a non-unitary transformation which alters the superposition state of the skyrmions, the topological invariant ${N}_\mathrm{skyr}$ remains intact. This contrasts with conventional polarization-based storage, where efficiency imbalance or phase errors severely degrade reconstruction fidelity. The resilience of skyrmionic structures highlights their potential as information carriers compatible with imperfect photonic memory platforms. Based on this mechanism, we implement a topologically protected optical memory. The experimental setup and results are described in the following section.

\paragraph{\textbf{Experimental setup:}}

\begin{figure*}[!t]
\centerline{\includegraphics[width=\linewidth]{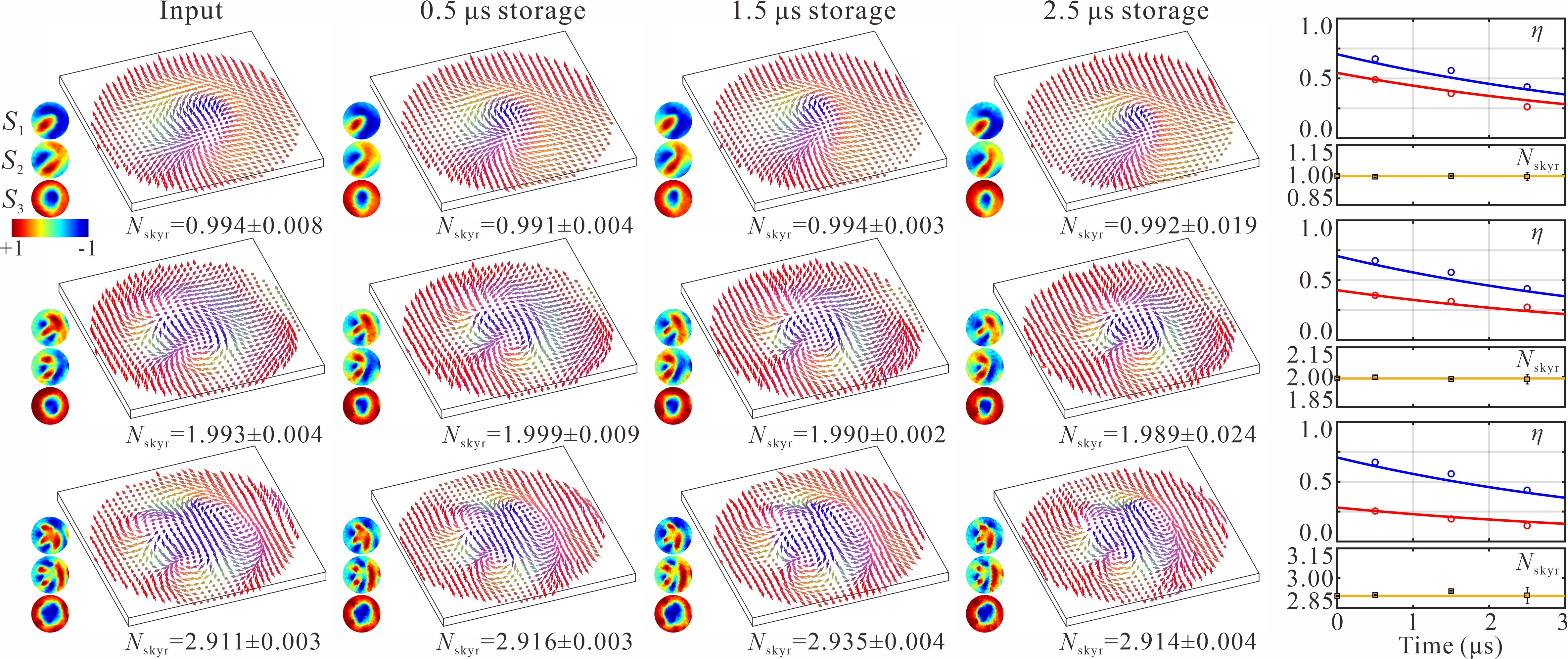}}
\caption{Left: The measured topological textures of input and output optical skyrmions with ${N}_\mathrm{skyr}$=1, 2 and 3 (from top to bottom). The experimentally calculated ${N}_\mathrm{skyr}$ are presented with respect to each sub-figure, with error bars representing the standard deviation of five runs. The insets on the side are the normalized local Stokes parameters. Right: Comparison between experimental data and theoretical simulation. The blue (red) solid lines represent the relationship of storage efficiency of the $\rm{LG}^{0}_{0}$ ($\rm{LG}^{\it{l}}_{0}$) mode against storage time, while the blue (red) circles represent the experimental data of the respective modes. The yellow solid line represents the measured input ${N}_\mathrm{skyr}$, and the data points represent the experimental data after storage.}
\label{fig3}
\end{figure*}

Various techniques have been developed for generating paraxial optical skyrmions \cite{shen2022generation,kerridge2024optical,lin2024chip,hakobyan2025q}. Here, we use a configuration comprising a segmented-screen spatial light modulator and an interferometer, as detailed in the Supplemental Material. Our simplified experimental setup for optical storage is illustrated in Fig. \ref{fig2}(a). The generated optical skyrmions are then stored in a cigar-shaped cloud of cold $^{87}$Rb atoms, with 3 cm length and 1.5 mm diameter, prepared within a two-dimensional magneto-optical trap (MOT) \cite{yang2025efficientOAM}. The relevant atomic $\Lambda$-system for the reversible mapping is based on three Zeeman sublevels of the $D_1$ line. Before storage, atoms are prepared in $|F=1,m_F=+1\rangle$ to maximize the absorption of the probe beams. Dual-path storage is realized by converting the optical skyrmion into a superposition of separate paths, as given in Eq.~(\ref{eq3}), where the usual orthogonal circular polarizations are unified to ${\boldsymbol\sigma}_+$ and the entanglement is instead encoded in the paths. To achieve this, a quarter-wave plate (Q) and a polarizing beam displacer (BD) separate the spatial modes $\rm{LG}^{0}_ {0}$ and $\rm{LG}^{\it{l}}_{0}$. Additional polarization operations allow us to obtain the desired polarization state. We also calibrate and optimize our system for equal mode profiles to ensure balanced OD in each path before storage, as detailed in the Supplemental Material.

During the optical storage process, the temporal waveform of the probe pulse is tailored to a Gaussian shape with $0.5\,\mu$s duration, and the control beam is adiabatically switched off once the probe pulse fully enters the atomic ensemble, allowing the optical states to be coherently transferred into a collective spin excitation, as shown in the top left inset of Fig.~\ref{fig2}(a). Upon reactivation of the control beam after a specific storage time, both components of the probe beam in the two paths are retrieved with their preserved spatial phase and intensity distribution, facilitated by a collective enhancement effect \cite{veissier2013reversible,ding2013single,nicolas2014quantum}. The temporal waveforms of both paths are obtained using photodetectors (not shown) to assess the storage efficiency. Lastly, by employing optical elements that unitarily reverse the decomposition process of the optical skyrmion, the retrieved skyrmion is re-assembled. Its spatial profile is subsequently reconstructed and imaged on an intensified charge-coupled device (ICCD).
%with a response time of $2\,$ns. 

\paragraph{\textbf{Experimental results:}}

Figure~\ref{fig2}(b) shows the temporal waveforms and storage efficiencies for the different modes. As expected, the efficiency is highest for the $\rm{LG}^{0}_{0}$ mode and decreases for higher-order modes (larger $l$) due to reduced light-atom overlap. The efficiency in each path decays with storage time due to spin-wave decoherence, which has been shown to substantially degrade the fidelity of stored information \cite{chen2021phase,wang2021efficient,ye2022long}. As a first step, we perform fully spatially resolved Stokes tomography on the states retrieved from the atomic memory. The Stokes parameters are recorded to reconstruct the spatial polarization profiles, as shown in the Supplemental Material. The resulting Stokes vector profiles are present in Fig.~\ref{fig3}.

We calculate ${N}_\mathrm{skyr}$ of the input mode in the absence of the MOT, as well as the corresponding retrieved results after storing $0.5\,\mu$s, $1.5\,\mu$s, and $2.5\,\mu$s. The experimental results show a minor deviation of ${N}_\mathrm{skyr}$ from their input values, and demonstrate the topological robustness of optical skyrmions under decoherence and the imbalanced storage efficiency between the paths, underscoring the intrinsic topological protection of skyrmionic configurations. Notably, slight deviations could be attributed to several factors, such as the imaging system, the background noise of our ICCD image and the fluorescence noise of the atoms, yet the overall trend confirms resilience of ${N}_\mathrm{skyr}$ against decoherence over practical storage timescales. This contrasts with the decay of concurrence observed in vector beams under similar perturbations, highlighting the superior topological protection afforded by the skyrmion number, as detailed in the Supplemental Material. During storage, the optical state is mapped into a stationary atomic spin wave, as a result, there is no free-space propagation and hence no accumulation of relative Gouy phase \cite{yao2011orbital}. As a result, the retrieved skyrmion retains the original topological texture without changing to a different skyrmion configuration. On the right side of Fig.~\ref{fig3}, we also present a comparison between the theoretical simulation and the experimental results, as detailed in the Supplemental Material.

\begin{figure}[!t]
\centerline{\includegraphics[width=\linewidth]{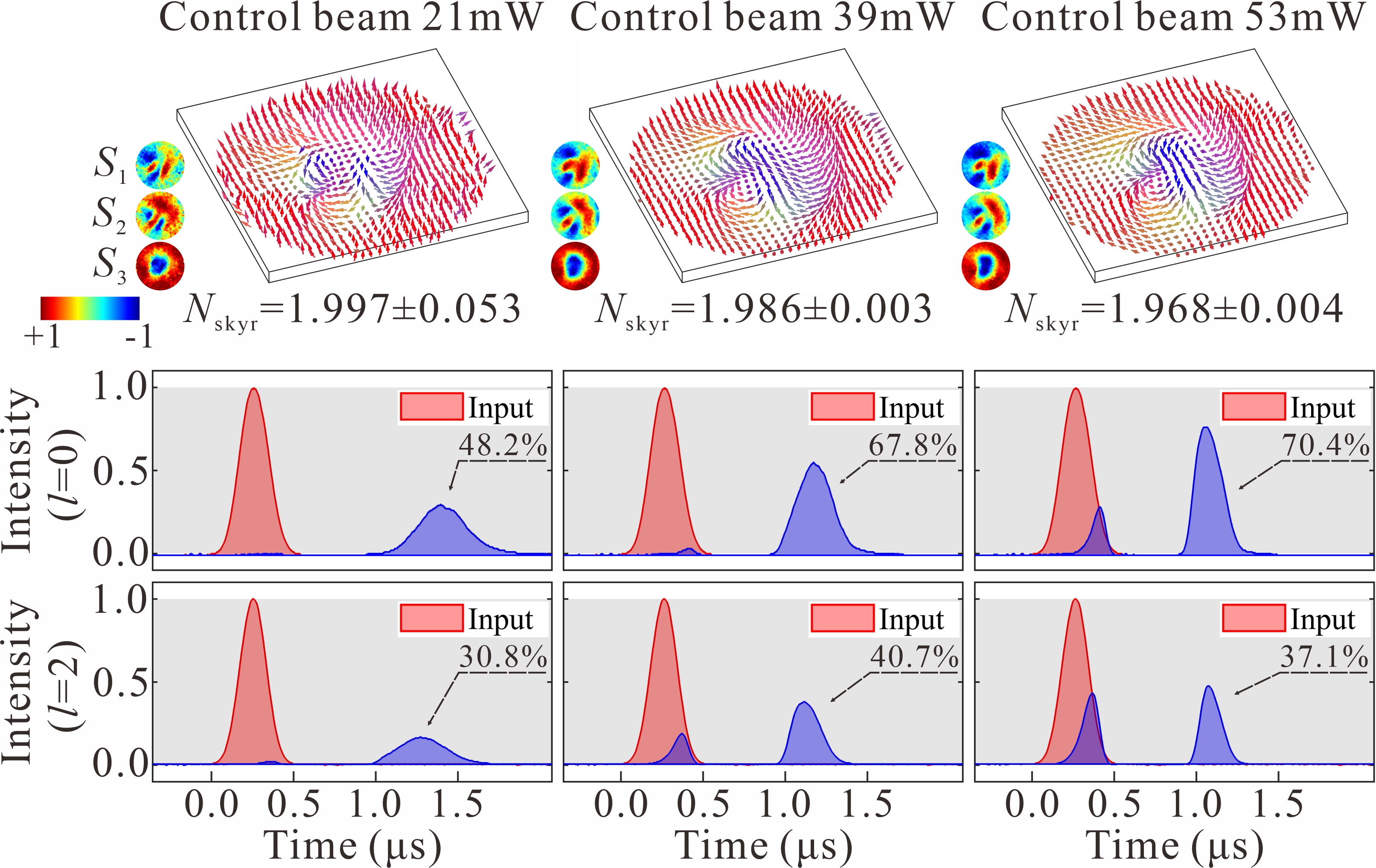}}
\caption{Experimental results of storage under different control beam power, for an optical skyrmion with ${N}_\mathrm{skyr}$=2 and a storage time of $0.5\,\mu$s. Top: The measured topological textures of input optical skyrmions with ${N}_\mathrm{skyr}$=2. The normalized local Stokes parameters and calculated ${N}_\mathrm{skyr}$ are on the left and below respectively. The error bars represent the standard deviation of five runs. Bottom: The input and retrieved temporal waveforms of the input $\rm{LG}^{0}_{0}$ and $\rm{LG}^{2}_{0}$ modes. Additionally, the storage efficiencies are calculated and presented. The non-stored remnants of the pulses are the leakage.}
\label{fig4}
\end{figure}

We also study the impact of the control beam power on the stored optical skyrmions, illustrated by an example with ${N}_\mathrm{skyr}$=2 and a storage time of $0.5\,\mu$s, as shown in Fig.~\ref{fig4}. Changes in control beam intensity modify the medium’s refractive response, which, owing to the differing spatial mode profiles of the two paths, results in distinct EIT responses. This alters the spatial polarization of the retrieved optical skyrmion, as detailed in the Supplemental Material. As the control beam power increases, the stored skyrmion number decreases slightly, but the topological texture remains clear and intact, highlighting its inherent stability. Under the strong coupling regime, the loss in each path decreases and the delay between the paths shortens \cite{lukin2001controlling}, while the imbalance in storage efficiency increases. The maximum power of 53 mW ($\approx$48.3 MHz) used in our experiments represents a practical upper bound set by our system constraints, and it also approaches the theoretical limit of the optimal control strength, as detailed in the Supplemental Material. In theory, ${N}_\mathrm{skyr}$ remains quantized until the storage process completely fails for one path, beyond which the superposition is no longer coherent. This perturbation, while causing changes in local ellipticity, polarization rotations, and minor variations in skyrmion number, does not disrupt the overall topological structure \cite{ye2025atlas}. We have therefore demonstrated that even large perturbations to the storage systems,  by changing the control beam power by more than 100\%, do not significantly perturb the topological structure of the retrieved skyrmion.

\paragraph{\textbf{Conclusions:}}
In summary, we have presented the first experimental demonstration that the topological characteristics of an optical skyrmion are preserved during coherent storage and retrieval in an optical memory. By using a dual-path EIT scheme, we directly demonstrated that the skyrmion number remains robust against experimental imperfections. This resilience is not merely the conservation of a classical number, but rather a direct manifestation of the topological protection inherent to the skyrmionic structure, which prevents the stored state from degenerating into a trivial configuration. Our work thus provides critical validation that the topological protection of skyrmions is not compromised by the complex light-matter interactions and decoherence processes central to quantum memories.

Our primary focus in this work is to establish a proof-of-principle demonstration of skyrmion storage and to investigate the preservation of its topological invariant. Further work could optimize the memory performance itself or investigate the stability of skyrmion textures under varying control beam parameters. In particular, exploring the potential disruption mechanisms at high-power levels could provide valuable insights into the robustness limits of topological optical memory.

Looking forward, the significance of this result extends beyond the storage of a single topological charge. It establishes that the fixed skyrmion number defines a topologically protected subspace within which quantum information can be encoded. Future applications can exploit this stability by encoding qubits in superpositions of states residing within this subspace, for instance, using different azimuth or radial indices while maintaining a fixed total skyrmion number \cite{zeissler2020diameter,wang2025generation,zeng2025tailoring}, or even in superpositions of distinct skyrmion numbers themselves \cite{psaroudaki2021skyrmion,psaroudaki2023skyrmion,petrovic2025colloquium}. Our findings thereby open a pathway toward robust quantum photonic technologies by combining topological photonics and quantum memory. The demonstrated resilience against realistic noise in EIT-based storage marks a foundational step to leverage topological invariance as a resource for protecting quantum information \cite{ornelas2024non,de2025quantum,ma2025nanophotonic,ornelas2025topological}.

%In summary, we have experimentally demonstrated optical skyrmion storage on a microsecond timescale and confirmed the preservation of their original topological characteristics after storage. Our procedure relies on the conversion of the skyrmion into a delocalized skyrmion, allowing us to facilitate dual-path EIT storage, where each path encodes orthogonal spatial modes. We validated that the skyrmion number of the retrieved optical skyrmions maintains robust even under conditions of imbalanced storage efficiencies and strong perturbations induced by the control beam. This invariance stems from the intrinsic topological protection inherent to skyrmionic configurations, which ensures the conservation of their global topological number against atomic spin wave decoherence. The storage scheme here may be generalized to other topological quasiparticles of light, such as optical hopfion or optical skyrmion lattices, as robust information carriers in photonic memory devices, leveraging their topologically protected states and invariance for both classical and quantum communication.

\begin{acknowledgments}

\section{acknowledgment}
This work was supported by National Natural Science Foundation of China (12404390, 12404406, 12104358, 12304406 and 92476105); Postdoctoral Fellowship Program of China Postdoctoral Science Foundation (GZC20232118); Shaanxi Province postdoctoral Science Foundation (2023BSHEDZZ23); the Fundamental Research Funds for the Central Universities (xzy012023042) and Huzhou Natural Science Foundation (2024YZ56). S. F.-A. acknowledges support through the QuantERA II Programme, with funding received via the EU H2020 research and innovation programme under Grant No. 101017733 and associated support from EPSRC under Grant No. EP/Z000513/1 (V-MAG), as well as support via the COST Action CA23134 (polytopo) funded by the European Union. C. M. C. wishes to acknowledge support from Early Career Leverhulme Fellowship (ECF-2023-099). Y. Shen acknowledges support through the Singapore Ministry of Education (MOE) AcRF Tier 1 grants (RG157/23 \& RT11/23) and the Singapore Agency for Science, Technology and Research (A*STAR) MTC Individual Research Grants (M24N7c0080).
\end{acknowledgments}

\section{References}
\bibliography{sample}
\end{document}